\documentclass[a4paper,twocolumn,superscriptaddress,prx, aps, floatfix ]{revtex4-2}
\usepackage{latexsym,amsmath}
\usepackage{amsbsy}
\usepackage[pdftex]{graphicx}
\usepackage{amssymb}
\usepackage{epstopdf}
\usepackage{verbatim}
\usepackage[usenames]{color}
\usepackage{capt-of} % Used for table caption
\usepackage{float}
\usepackage{ esint }
\usepackage{hyperref}
\usepackage{graphicx}
\usepackage[normalem]{ulem}

\usepackage{color}

\begin{document}

\title{Emergence of geometric Turing patterns in complex networks}

\author{Jasper van der Kolk}
\email{jasper.vanderkolk@ub.edu}
\affiliation{Departament de F\'isica de la Mat\`eria Condensada, Universitat de Barcelona, Mart\'i i Franqu\`es 1, E-08028 Barcelona, Spain}
\affiliation{Universitat de Barcelona Institute of Complex Systems (UBICS), Barcelona, Spain}

\author{Guillermo \surname{Garc\'ia-P\'erez}}
%\email{marian.serrano@ub.edu}
\affiliation{Algorithmiq Ltd, Kanavakatu 3C, FI-00160 Helsinki, Finland}

\author{Nikos E.~\surname{Kouvaris}}
%\email{marian.serrano@ub.edu}
\affiliation{Dribia Data Research S.~L., Barcelona, Spain}

\author{\\M. \'Angeles \surname{Serrano}}
\email{marian.serrano@ub.edu}
\affiliation{Departament de F\'isica de la Mat\`eria Condensada, Universitat de Barcelona, Mart\'i i Franqu\`es 1, E-08028 Barcelona, Spain}
\affiliation{Universitat de Barcelona Institute of Complex Systems (UBICS), Barcelona, Spain}
\affiliation{Instituci\'o Catalana de Recerca i Estudis Avan\c{c}ats (ICREA), Passeig Llu\'is Companys 23, E-08010 Barcelona, Spain}

\author{Mari\'an Bogu\~n\'a}
\email{marian.boguna@ub.edu}
\affiliation{Departament de F\'isica de la Mat\`eria Condensada, Universitat de Barcelona, Mart\'i i Franqu\`es 1, E-08028 Barcelona, Spain}
\affiliation{Universitat de Barcelona Institute of Complex Systems (UBICS), Barcelona, Spain}

\begin{abstract}
Turing patterns, arising from the interplay between competing species of diffusive particles, has long been an important concept for describing non-equilibrium self-organization in nature, and has been extensively investigated in many chemical and biological systems. Historically, these patterns have been studied in extended systems and lattices. Recently, the Turing instability was found to produce topological patterns in networks with scale-free degree distributions and the small world property, although with an apparent absence of geometric organization. While hints of explicitly geometric patterns in simple network models (e.g Watts-Strogatz) have been found, the question of the exact nature and morphology of geometric Turing patterns in heterogeneous complex networks remains unresolved. In this work, we study the Turing instability in the framework of geometric random graph models, where the network topology is explained by an underlying geometric space. We demonstrate that not only can geometric patterns be observed, their wavelength can also be estimated by studying the eigenvectors of the annealed graph Laplacian. Finally, we show that Turing patterns can be found in geometric embeddings of real networks. These results indicate that there is a profound connection between the function of a network and its hidden geometry, even when the associated dynamical processes are exclusively determined by the network topology.
\end{abstract}

\maketitle

\tableofcontents

\section{Introduction}

In his 1952 seminal paper~\cite{Turing1952}, Alan Turing described how competing species of diffusive particles can cause a wide variety of patterns, starting from a spatially homogeneous state. Ever since, reaction-diffusion processes have been shown to host a wide variety of self-organizational behavior~\cite{gierer1972theory,meinhardt1982models,Cross:1993wq,koch1994biological,Zhabotinsky1995,cross2009pattern,kondo2010reaction,walgraef2012spatio}. Historically, these behaviors have mostly been studied on regular lattices or in continuous media, but there has long been evidence for their existence also in systems of interconnected elements with complex structures. For instance, in the framework of embryonic morphogenesis, which was Turing's original motivation for the development of his theory of pattern formation, it has been argued that embryos should be thought of as a multicellular network rather than a continuous medium~\cite{Othmer1971}. However, complex network architecture makes the investigation of dynamical processes difficult, and studies have often been restricted to small networks~\cite{Mikhailov:2010,Horsthemke2004,Moore2005}. 

In the context of network science, Nakao and Mikhailov made a step forward and demonstrated the existence of the Turing instability in systems of activator-inhibitor species diffusing in large random graphs~\cite{Mikhailov:2010}. In that work, the instability and the corresponding emerging patterns are intimately related to the degree heterogeneity usually present in real networks and are, thus, purely topological in nature. While hints of explicitly geometric patterns in simple network models have been found in subsequent works~\cite{Asllani:2014cr,CENCETTI2020109707,Hutt:2022gf}, no comprehensive study of this phenomenon has been undertaken. Most importantly, the network models described in these studies are not suitable to describe real systems. In fact, in real complex networks, geometric domains, so typical of Turing patterns in lattices and continuous media, were not observed. This is a consequence of the apparent lack of geometric structure in complex small-world networks, causing topological distances between nodes --measured by the shortest path lengths on the graph-- to collapse around its average value, scaling logarithmically (or slower) with the system size. 

This paradigm has changed recently with the development of network geometry~\cite{Boguna2021}. Indeed, a latent metric space with hyperbolic geometry underlying real complex networks provides the simplest explanation to many of their observed topological properties, including heterogeneous degree distributions~\cite{serrano2008similarity,krioukov2010hyperbolic,gugelmann2012random}, clustering~\cite{krioukov2010hyperbolic,gugelmann2012random,candellero2016clustering,Fountoulakis2021,Jasper_2022}, small-worldness~\cite{abdullah2017typical,friedrich2018diameter,muller2019diameter}, and percolation~\cite{serrano2011percolation,fountoulakis2018law}, spectral~\cite{kiwi2018spectral} and self-similarity properties~\cite{serrano2008similarity,garcia-perez2018multiscale,Zheng:2021aa}. These models have been extended to growing networks~\cite{papadopoulos2012popularity}, weighted networks~\cite{allard2017geometric}, multilayer networks~\cite{kleineberg2016hidden,Kleineberg2017}, networks with community structure~\cite{zuev2015emergence,garcia-perez:2018aa,muscoloni2018nonuniform} and they are also the basis for defining a renormalization group for complex networks~\cite{garcia-perez2018multiscale,Zheng:2021aa}. In this paper, we show that the Turing instability triggers the emergence of purely geometric patterns that become evident in the latent space of real complex networks. We analyze this phenomenon within the framework of network geometry and find the relation between the parameters of the dynamical model and the general properties of the network, such as the congruency of the network with the underlying network space and the level of heterogeneity of its degree distribution.

\section{Background}

The Turing instability arises as a consequence of the different diffusivity rates between activator and inhibitor species. Starting from a homogeneous concentration of particles, upon small perturbations, the system evolves towards a stable state with regions of high concentration of particles, coexisting with regions of low concentration. When the dynamics take place on a metric space, these inhomogeneities give rise to geometric patterns. It is important to note that the first assumption of this framework is that diffusion alone, without the activation-inhibition mechanism, should lead to a homogeneous distribution of particles. 

In continuous media, diffusion is modeled by brownian particles and, in degree regular lattices, by standard random walks. Diffusion in complex networks is, however, different. Indeed, a simple random walk process on a heterogeneous network leads to a steady state where the concentration of particles is proportional to the degree of each node and, so, highly heterogeneous. Thus, before adding the activation-inhibition dynamics, we have to carefully define diffusion on networks to ensure that a homogeneous concentration of species is a fixed point of the dynamics. In the Appendix~\ref{appendix:diffusion} we show that this can be achieved by a continuous time random walk with jump rates proportional to the current node's degree. Then, the evolution equation for the average concentration of particles at a given node $i$, $u_i(t)$ can be written as
\begin{equation}
\label{eq:difussion}
\frac{du_i(t)}{dt}=-\epsilon \sum_{j=1}^N L_{ij}u_j(t),
\end{equation}
where $\epsilon$ is the diffusion coefficient, $N$ the number of nodes, and $L_{ij}$ the Laplacian matrix defined as in Ref.~\cite{mieghem_2010}:
\begin{equation}
L_{ij}=k_j \delta_{ij}-a_{ij}.
\end{equation}
In this last equation $a_{ij}$ is the adjacency matrix of the network, that hereafter is assumed to describe a single connected component.

\subsection{Properties of the Laplacian matrix}

As we shall see in the next sections, the Laplacian matrix plays a central role in the Turing instability. The properties of the Laplacian matrix have been extensively studied~\cite{mieghem_2010} and some of them will be used in our work. The Laplacian is a real symmetric matrix with real eigenvalues $\Lambda_1=0 \le \Lambda_2 \le \cdots \le \Lambda_N \le 2k_c$, where $k_c$ is the maximum degree of the network. The average of all eigenvalues is equal to the network average degree, that is,
\begin{equation}
\langle \Lambda \rangle=\langle k \rangle
\label{averageLambda}
\end{equation}
and the second moment is
\begin{equation}
\langle \Lambda^2 \rangle= \langle k^2 \rangle+\langle k \rangle.
\end{equation}
These results suggest that eigenvalues are strongly correlated to the sequence of degrees in the network. Besides, as we show in App.~\ref{appendix_chegger}, in geometric random graphs models, bounds on the isoperimetric (or Cheeger) constant~\cite{Polya:1951aa,Cheeger:1969aa} imply that the smallest non-null eigenvalue --the spectral gap-- of the Laplacian approaches zero in the thermodynamic limit. If, in addition, the degree distribution is unbounded, we can assume that there exists an eigenvalue close to any point in $\mathbb{R}^+\cup \{0\}$. For this reason, in this paper we are not concerned by the actual value of the eigenvalues but by the form of the eigenvector given a particular eigenvalue $\Lambda$. We shall see, however, that the most relevant eigenvalues for the Turing instability are the small ones, that is, those satisfying $\Lambda \ll \langle k \rangle$.

In turn, eigenvectors of the Laplacian matrix $\{\vec{\phi}^\alpha=(\phi_1^\alpha,\phi_2^\alpha,\cdots,\phi_N^\alpha) \; ; \; \alpha=1,\cdots,N \}$ form a complete orthonormal basis of $\mathbb{R}^N$ and, thus, can be used as a basis to express the dynamic state of the system. The eigenvector of zero eigenvalue is constant and the rest of eigenvectors satisfy the condition
\begin{equation}
\sum_{i=1}^N \phi_i^\alpha=0,
\label{vanishcomponents}
\end{equation}
so that the components of each eigenvector must necessarily balance positive and negative values. As we see next, this property turns out to be critical to determine the form of the Turing patterns.

\subsection{Activation-inhibition dynamics}

In the diffusion process described by Eq.~\eqref{eq:difussion}, particles do not interact when they meet in the same node. However, in realistic settings, particles may undergo all sorts of reaction processes. In particular, the Turing instability arises when two different species, $U$ and $V$, interact upon meeting at the same node, under an activation-inhibition process. In particular, within each node, species $U$ undergoes an autocatalytic process and, simultaneously, favors the increase of the population of species $V$. At the same time, species $V$ inhibits the growth of species $U$ even though it cannot survive without it. When the average number of particles per node is large, we can neglect fluctuations in the number of particles and work under the mean field approximation. Let $u_i(t)$ and $v_i(t)$ be the average number of particles at node $i$ of species $U$ and $V$, respectively. Their evolution equations are given by
\begin{eqnarray}
\frac{du_i(t)}{dt} & = & f\left(u_i (t), v_i (t)\right) - \epsilon \sum \limits_{j=1}^{N} L_{ij} u_j (t) \label{eq:master1}\\
\frac{dv_i(t)}{dt} & = & g\left(u_i (t), v_i (t)\right) - \sigma \epsilon \sum \limits_{j=1}^{N} L_{ij} v_j (t),\label{eq:master2}
\end{eqnarray}
where functions $f(u,v)$ and $g(u,v)$ represent the activation-inhibition dynamics within the nodes and $\epsilon$ and $\sigma \epsilon$ are the activator and inhibitor diffusion coefficients, respectively. We further assume that the system has a stable homogeneous stationary state, that is $u_i^{st}=\bar{u}$ and $v_i^{st}=\bar{v}$ for which $f(\bar{u}, \bar{v}) = 0$ and $g(\bar{u}, \bar{v}) = 0$ so that $(\bar{u}, \bar{v})$ is a stable fixed point of the dynamics. The activation-inhibition dynamics and its stability condition impose constraints on functions $f(u,v)$ and $g(u,v)$. In particular, its first derivatives $f_u \equiv \partial_u f(\bar{u}, \bar{v})$, $f_v \equiv \partial_v f(\bar{u}, \bar{v})$, $g_u \equiv \partial_u g(\bar{u}, \bar{v})$, and $g_v \equiv \partial_v g(\bar{u}, \bar{v})$ must satisfy $f_u>0,f_v<0$, $g_u>0,g_v<0$ and $f_u+g_v<0$, $f_ug_v-f_vg_u>0$ (see~\cite{Mikhailov:2010} for further details).

The Turing instability arises when, due to the differences in the diffusion coefficients of the two species, such fixed point becomes unstable so that any small perturbation drives the system away from it. Interestingly, the conditions for this to take place only depend on $f_u,f_v,g_u,g_v$ and the ratio between the diffusion coefficients of the two species $\sigma$. The behavior of the system near the fixed point can be expanded in terms of the eigenvectors of $L_{ij}$ as~\cite{Mikhailov:2010}
\begin{equation}
u_i(t)-\bar{u} \approx \sum_{\alpha=1}^N c_\alpha e^{\lambda_\alpha t} \phi_i^\alpha,
\end{equation}
where $c_\alpha$ and $\lambda_\alpha$ are some constants that depend on the parameters of the model and on the eigenvalue $\Lambda_\alpha$ (similarly for species $V$). If $\lambda_\alpha<0 \; ; \; \forall \Lambda_\alpha$, then perturbations around the fixed point are absorbed exponentially fast, so that the fixed point is stable. However, if there exists at least one $\Lambda_\alpha$ such that $\lambda_\alpha>0$, then the fixed point becomes unstable. In this case, those nodes with $c_\alpha \phi_i^{\alpha}>0$ will increase the concentration of species $U$ whereas those with negative value will decrease it. Since the components of all eigenvectors of the Laplacian matrix must have positive and negative values summing up to zero, and each component is associated to a given node of the network, the dynamics will necessarily evolve towards a state with some of the nodes containing a high concentration of species $U$ and the rest a low concentration.

In~\cite{Mikhailov:2010} it was shown that there exist unstable eigenvalues (i.e $\lambda_\alpha>0$) whenever
\begin{equation}
\sigma \ge \sigma_c=\frac{f_v g_u}{f_u^2} \left[ \frac{f_u g_v}{f_v g_u}-2-2\sqrt{1-\frac{f_u g_v}{f_v g_u}}\right].
\label{eq:sigmac}
\end{equation}
In this regime, all eigenvalues that lay within the interval $\Lambda \in [\Lambda_-,\Lambda_+]$ with
\begin{equation}
\Lambda_\pm = \frac{1}{2\sigma \epsilon}[\sigma f_u+g_v \pm \sqrt{(\sigma f_u-g_v)^2+4\sigma f_vg_u}]
\label{eq:interval}
\end{equation}
become unstable, with the most unstable one being
\begin{equation}
\Lambda_{\text{max}}=\frac{1}{1-\sigma}\left(f_u-g_v-(1+\sigma)\sqrt{\frac{-f_v g_u}{\sigma}} \right).
\end{equation}
The functional dependence of Eqs.~\eqref{eq:sigmac} and~\eqref{eq:interval} implies that, for a given set of parameters $f_u,f_v,g_u,g_v$ and for one particular eigenvalue $\Lambda_\alpha$, it is always possible to select parameters $\epsilon$ and $\sigma$ such that $\Lambda_\alpha \in [\Lambda_-,\Lambda_+]$. In the case that $\Lambda^+\approx\Lambda^-\approx\Lambda_\alpha$ one can assume that the differences in the concentration of species in the nodes are encoded in the eigenvector $\vec{\phi}^\alpha$.

Interestingly, these results only depend on the first derivatives of functions $f(u,v)$ and $g(u,v)$ coupled to the diffusion coefficients. Therefore, there exists a whole class of systems where such instability may emerge, from chemical reactions~\cite{prigogine1968symmetry,castets1990experimental,ouyang1991transition} or biological morphogenesis~\cite{meinhardt1982models,harris2005molecular} to ecosystems~\cite{Segel:1976aa,MIMURA1978249,maron1997spatial,Gibert:2019aa} and game theory applied to ecological systems~\cite{Fahimipour:2022aa,Brechtel:2018aa}. In ecology, one such model is the Mimura-Murray model~\cite{MIMURA1978249}. Here, $u$ and $v$ represent prey and predator densities, respectively, and the dynamics is governed by the functions
\begin{eqnarray}
f(u, v) & = & \left( \frac{a + bu - u^2}{c} - v \right) u, \label{eq:MimuraMurray1}\\
g(u,v) & = & \left( u - \left( 1 + dv \right) \right) v.
\label{eq:MimuraMurray2}
\end{eqnarray}
From Eqs.~\eqref{eq:master1},~\eqref{eq:master2},~\eqref{eq:MimuraMurray1}, and \eqref{eq:MimuraMurray2}, we see that in the absence of preys, predators go extinct and in the absence of predators, preys attain a constant population. In general, there is a fixed point with positive densities of preys and predators. In this paper, we use this dynamics as a case study (technical details are given in the Appendix~\ref{appendix_dynamics}).

\subsection{A geometric description of complex networks. The $\mathbb{S}^1/\mathbb{H}^2$ model}
\label{s1model}

As we have seen, the shape of emerging patterns is dictated by the arrangements of the signs in the components of the eigenvectors of the Laplacian matrix. In continuous media or regular lattices, these signs are arranged in the metric space in an alternating fashion, so that when the parameters of the dynamics are tuned to select these eigenvectors as unstable, the concentration of species also follows the same geometric patterns. In random graphs models, like the configuration model~\cite{bender1978asymptotic,bollobas1980probabilistic,molloy1995critical,molloy1998size} or the Bar\'abasi-Albert model~\cite{barabasi1999emergence}, the situation is different. For these type of models, there is no associated metric space and the most distinctive feature of nodes is the degree. In this case, the Turing instability leads to patterns that are purely topological, with high or low concentration of species depending on nodes' degrees~\cite{Mikhailov:2010}. Real complex networks, however, are better described by geometric random graph models. In such models, nodes are embedded in a metric space and the connection probability depends on the distances among nodes in this metric space. This approach has lead to the development of the field of network geometry~\cite{Boguna2021}, giving rise to the most comprehensive description of real complex networks. 

To generate geometric networks, we use the $\mathbb{S}^1/\mathbb{H}^2$ model~\cite{serrano2008similarity,krioukov2010hyperbolic,serrano_boguna_2022}, otherwise known as the geometric soft configuration model~\cite{Boguna:2020fj}. The model combines a similarity dimension, encoded in a $1$-dimensional sphere, and a popularity dimension, quantified by nodes' degrees. In particular, each node is given a pair of hidden variables $\kappa_i \in [\kappa_0,\infty)$ and $\theta_i \in [0,2\pi)$. The former accounts for the $i$'th node's ensemble average degree, and can be generated from an arbitrary distribution $\rho(\kappa)$. The latter is the angular position of node $i$ on the circle, as an abstraction of its position in the similarity space. In the simplest version of the model, $\kappa$ and $\theta$ are statistically independent random variables and $\theta$ is homogeneously distributed. This final assumption, however, is not critical to our model and, in fact, in real systems the angular distribution is inhomogeneous, inducing the emergence of geometric communities~\cite{boguna2010sustaining,serrano2012uncovering,zuev2015emergence,garcia-perez:2018aa,muscoloni2018nonuniform}. The model is fully determined by the following connection probability between two nodes $i$ and $j$
\begin{equation}
p_{ij}=\frac{1}{1+\left[ \frac{x_{ij}}{\mu \kappa_i \kappa_j}\right]^{\beta}},
\label{pijS1}
\end{equation}
where $x_{ij}$ is the distance between the nodes in the circle of radius $R=N/2\pi$, $\beta>1$ is the inverse of the temperature of the ensemble~\cite{krioukov2010hyperbolic,Boguna:2020fj,Jasper_2022}, and $\mu=\frac{\beta}{2\pi \langle k \rangle} \sin{\frac{\pi}{\beta}}$ a parameter fixing the average degree $\langle k \rangle$ of the network. While the model is defined for arbitrary distributions of hidden degrees, here we chose a power law distribution of the form $\rho(\kappa)=(\gamma-1) \kappa_0^{\gamma-1} \kappa^{-\gamma}$, with $\kappa \ge \kappa_0$. With this choice, the model becomes small-world whenever $1< \beta \le 2$, or $2<\gamma\le 3$, or both. Besides, degree heterogeneity is modulated by the exponent $\gamma$ and the limit $\gamma \rightarrow \infty$ is equivalent to a homogeneous distribution of hidden degrees $\rho(\kappa)=\delta(\kappa-\langle k \rangle)$ and  a Poisson degree distribution.

As it is defined, the $\mathbb{S}^1$ model represents the only maximally random ensemble of geometric graphs that are simultaneously sparse, clustered, small-world, with heterogeneous degree distribution, and with only structural degree-degree correlations~\cite{Boguna:2020fj}. In the regime $\beta>1$, the connection probability does not depend on the system size and, therefore, it generates networks with finite clustering in the thermodynamic limit. In this paper, we restrict ourselves to this regime of temperatures. However, the model has been extended to the case $\beta<1$, showing interesting quasi-geometric properties in the neighborhood of $\beta \lesssim1$ even though clustering vanishes for $\beta \le1$ in the thermodynamic limit~\cite{Jasper_2022}. It is worth mentioning that the model is defined in arbitrary dimensions. However, as shown in~\cite{Almagro:2021}, most real networks are well described by low dimensional spaces. Thus, to keep the analysis simple, we restrict ourselves to the one-dimensional case.

The $\mathbb{S}^1$ model is isomorphic to a purely geometric model in the hyperbolic plane of constant negative curvature $-1$, the so-called $\mathbb{H}^2$~\cite{krioukov2010hyperbolic} model. Upon mapping the hidden degree $\kappa$ to a radial coordinate $r$ as
\begin{equation}
r=R_{\mathbb{H}^2}-2 \ln{\frac{\kappa}{\kappa_0}},
\end{equation}
the connection probability Eq~\eqref{pijS1} becomes
\begin{equation}\label{eq:conn_prob}
p_{ij}= \frac{1}{1+e^{\frac{\beta}{2}(d_{\mathbb{H}^2,ij} - R_{\mathbb{H}^2})}},
\end{equation}
where $R_{\mathbb{H}^2}=2\ln{\frac{N}{\mu \pi \kappa_0^2}}$ is the radius of a hyperbolic disk and $d_{\mathbb{H}^2,ij}$ is the hyperbolic distance between nodes $i$ and $j$ given by the hyperbolic law of cosines,
\begin{equation}
\cosh{\left(d_{\mathbb{H}^2,ij}\right)}= \cosh{r_i} \cosh{r_j} - \sinh{r_i} \sinh{r_j} \cos{\Delta \theta_{ij}},
\end{equation}
where $\Delta \theta_{ij}$ is the angular separation between the nodes. Due to this isomorphism, we are free to work with one version or the other indistinctly. In this paper, we use the $\mathbb{S}^1$ model to perform the analytical calculations and the $\mathbb{H}^2$ model for visualization.

\section{The Mimura-Murray dynamics on geometric networks}

\begin{figure}[t]
\centering
\includegraphics[width=\columnwidth]{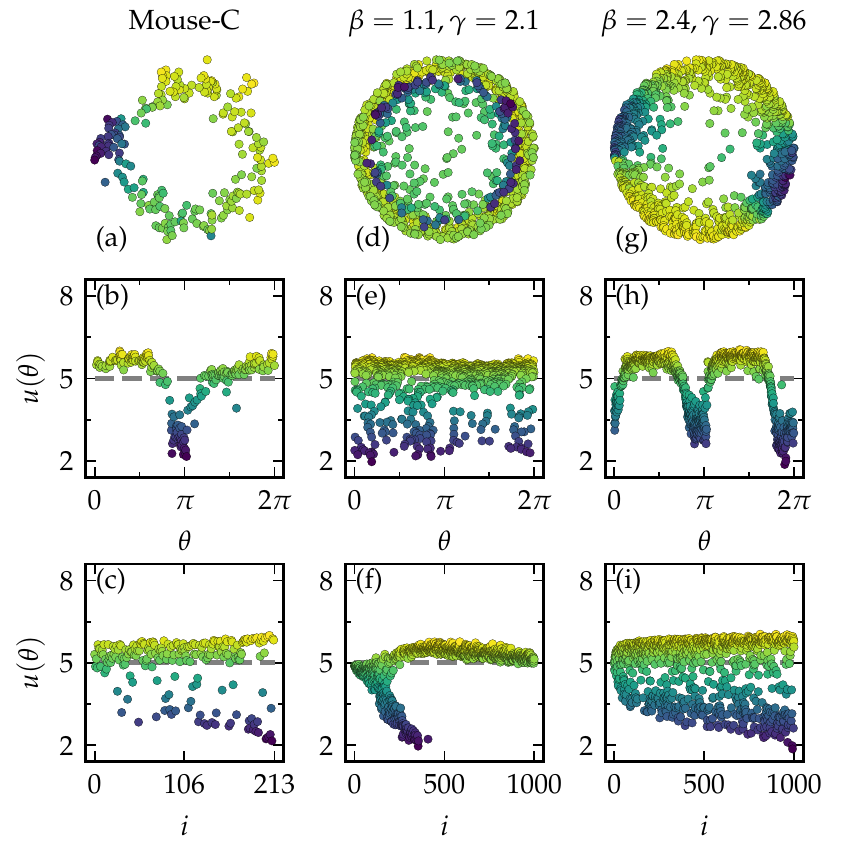}
\caption{Examples of Turing patterns in networks. The first, panels (a-c), is a real network representing the connectome of a mouse, the second, panels (d-f), a strongly heterogeneous $\mathbb{S}^1/\mathbb{H}^2$ network with parameters $(N, \langle k \rangle, \beta, \gamma) = (1000, 50, 1.1, 2.1)$ and the third, panels (g-i) a weakly heterogeneous $\mathbb{S}^1/\mathbb{H}^2$ network with parameters $(N, \langle k \rangle, \beta, \gamma) = (1000, 50, 2.4, 2.86)$. In all panels the colors correspond to the density of the activators in the final stationary state. In panels (a,d,g), nodes are located according to their coordinates in hyperbolic space. Panels (b,e,h) show the density of activators as a function of nodes' angular coordinates and in panels (c,f,i) the activators densities are plotted against nodes' degree ranking.}
\label{fig:results}
\end{figure}

To illustrate the role of geometry on the Turing instability, we run the Mimura-Murray dynamics on a real network, namely that of the connectome of a mouse described in~\cite{Oh2014} and whose properties are shown in Tab.~\ref{tab:RealNets}. Figure~\ref{fig:results}a shows the results on the concentration of species $U$ in the hyperbolic representation of the network (see Sec.~\ref{real:networks} for technical details). As can be seen, there is a clear pattern associated to the geometric organization of the network, with low concentrations of particles localized in the same angular position and high concentrations in the rest of the network. In panels \ref{fig:results}b and \ref{fig:results}c, we also represent the concentration as a function of the angular coordinate $\theta$ and the node index $i$, respectively. In the latter case the nodes are ordered from high to low degree. It is important to notice that the Mimura-Murray dynamics does not use any information from the underlying metric space, so that the observed geometric pattern is a highly non-trivial result. 

To shed light on this problem, we also run the dynamics on networks generated by the $\mathbb{S}^1/\mathbb{H}^2$ model with a heterogeneous degree distribution with exponent $\gamma$. Fig.~\ref{fig:results}d-i shows results of the dynamics on two different networks: one highly heterogeneous ($\gamma=2.1$) and weakly clustered and a second one weakly heterogeneous ($\gamma=2.84$) and highly clustered. In the strongly heterogeneous network shown in the figure, there are no patterns associated to the geometric character of the network. We do observe, however, topological patterns induced by degree, with high degree nodes holding a low concentration of species $U$. These are the type of patterns documented in~\cite{Mikhailov:2010}. Instead, in the case of the weakly heterogeneous network, we find a clear geometric pattern associated to the angular coordinates of nodes and no degree-related patterns, similar to the results found in the real network. These patterns emerge despite the fact that networks have the small-world property, so that diffusion induces flows of species between nodes that are far apart in the underlying metric space. These results suggest that some eigenvectors of the Laplacian of the $\mathbb{S}^1/\mathbb{H}^2$ model have a well defined periodic structure in the similarity space. Next, we develop a theoretical framework allowing us to calculate an approximation for the frequency of such eigenvectors.

\subsection{Eigenvectors of the $\mathbb{S}^1/\mathbb{H}^2$ Laplacian matrix in the annealed approximation}

Species whose dynamics are governed by the Mimura-Murray model diffuse within a network with a disordered but quenched structure; a single realization of the network ensemble defined by the $\mathbb{S}^1/\mathbb{H}^2$ model for a given sequence of hidden variables $(\kappa_i,\theta_i)$, $i=1,\cdots,N$. Unfortunately, for quenched networks it is not possible to get any analytical insight on the structure of the eigenvectors of the Laplacian matrix. To overcome this problem, we use the annealed approximation, which has been extensively used in the literature to tackle a wide variety of problems, from neural networks and opinion dynamics to epidemic spreading~\cite{derrida1986random,amari1992four,bastolla1996closing,luque1997phase,rohlf2002criticality,vilone2004solution,boguna2009langevin,guerra2010annealed,ferreira2011quasistationary}. In this approach, the network structure is resampled from the ensemble at a rate faster than the diffusion dynamics. This allows us to replace the adjacency matrix $a_{ij}$ by the connection probability $p_{ij}$ and consider the network not as a quenched system but as a dynamic one. With this approximation, the eigenvalue problem of the annealed Laplacian matrix can be written as
\begin{equation}
\sum_{j \ne i} \frac{\phi(\kappa_j,\theta_j)}{1+\left[ \frac{x_{ij}}{\mu \kappa_i \kappa_j}\right]^{\beta}}=\left[\sum_{j \ne i} \frac{1}{1+\left[ \frac{x_{ij}}{\mu \kappa_i \kappa_j}\right]^{\beta}}-\Lambda\right] \phi(\kappa_i,\theta_i),
\label{eigen}
\end{equation}
where $\Lambda$ is the eigenvalue and $\phi(\kappa_i,\theta_i)$ the component of the corresponding eigenvector of a node $i$ with hidden variables $(\kappa_i,\theta_i)$. Notice that, as in the case of the true Laplacian matrix, $\Lambda=0$ is an eigenvalue with constant eigenvector.

\begin{figure}[t]
\centering
\includegraphics[width=\columnwidth]{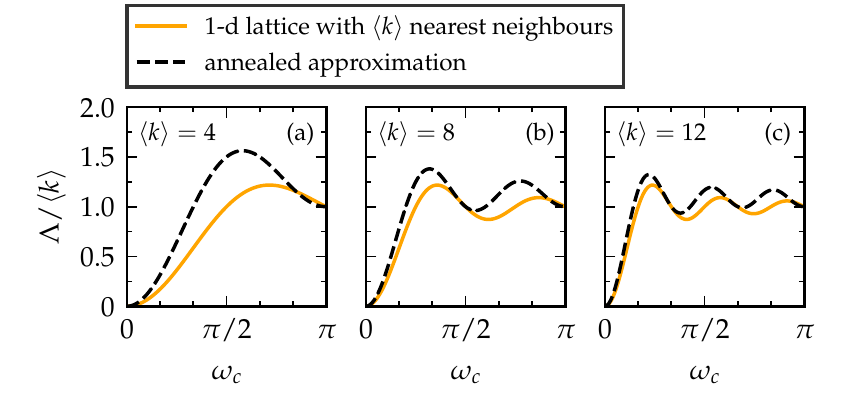}
\caption{\label{frequency1d}Dispersion relation for the annealed approximation and a degree-regular 1-dimensional lattice. Solid orange lines show exact results from Eq.~\eqref{lattice1d} and black dashed lines for the annealed approximation in Eq.~\eqref{modelbetainfinity}. Results are shown for average degrees $k=\langle k \rangle=4,8,$ and $12$.}
\end{figure}

In the thermodynamic limit, the curvature of the circle goes to zero and, thus, nodes become distributed in $\mathbb{R}^1$ according to a Poisson point process of density one. In this limit, the distance between two nodes can be evaluated as $x_{ij}=|x_i-x_j|$, where $x_i$ and $x_j$ are the positions of nodes $i$ and $j$ in $\mathbb{R}^1$. Finally, we take the continuum limit in Eq.~(\ref{eigen}) by replacing the sum over index $j$ by a double integral over nodes coordinates $(\kappa,x)$. In this way, Eq.~(\ref{eigen}) can be written as the following integral equation
\begin{equation}
\int_{\kappa_0}^\infty \rho(\kappa') d\kappa' \int_{-\infty}^{\infty} dx' \frac{\phi(\kappa',x')}{1+\left[ \frac{|x-x'|}{\mu \kappa \kappa'}\right]^{\beta}}=(\kappa-\Lambda) \phi(\kappa,x).
\label{eigen2}
\end{equation}
The spatial part of Eq.~\eqref{eigen2} has the form of a convolution integral. Thus, we can take advantage of the convolution theorem for Fourier transforms. By defining the Fourier transform of the eigenvector as
\begin{equation}
\hat{\phi}(\kappa,\omega) \equiv \int_{-\infty}^{\infty} e^{-i \omega x} \phi(\kappa,x)dx
\end{equation}
Eq.~(\ref{eigen2}) can be written in Fourier space as
\begin{equation}
\int_{\kappa_0}^\infty \frac{\kappa' \rho(\kappa')}{\langle k \rangle} \frac{\hat{\Psi}(\mu \kappa \kappa' \omega)}{\hat{\Psi}(0)} \hat{\phi}(\kappa',\omega) d\kappa'= \frac{\kappa-\Lambda}{\kappa} \hat{\phi}(\kappa,\omega),
\label{eigen3}
\end{equation}
where we have defined $\hat{\Psi}$ as 
\begin{equation}
\hat{\Psi}(z) \equiv \mathcal{F}\left[ \frac{1}{1+|x|^{\beta}}\right](z),
\end{equation}
where $ \mathcal{F}$ stands for Fourier transform. Notice that $\mu^{-1}=\langle k \rangle \hat{\Psi}(0)$. From Eq.~\eqref{eigen3}, it is easy to see that the integral over the spatial coordinate of non-trivial eigenvectors must vanish, that is, $\hat{\phi}(\kappa,0)=0$, similarly to the case of the quenched Laplacian matrix Eq.~\eqref{vanishcomponents}. Indeed, by setting $\omega=0$ in Eq.~\eqref{eigen3}, we conclude that $\hat{\phi}(\kappa,0)=0$ is the only solution when $\Lambda \ne 0$. Thus, the annealed approximation preserves the basic property of the Laplacian matrix given in Eq.~\eqref{vanishcomponents}.

\subsection{Homogeneous ensemble}

\begin{figure*}[t]
	\centering
	\includegraphics[width=\linewidth]{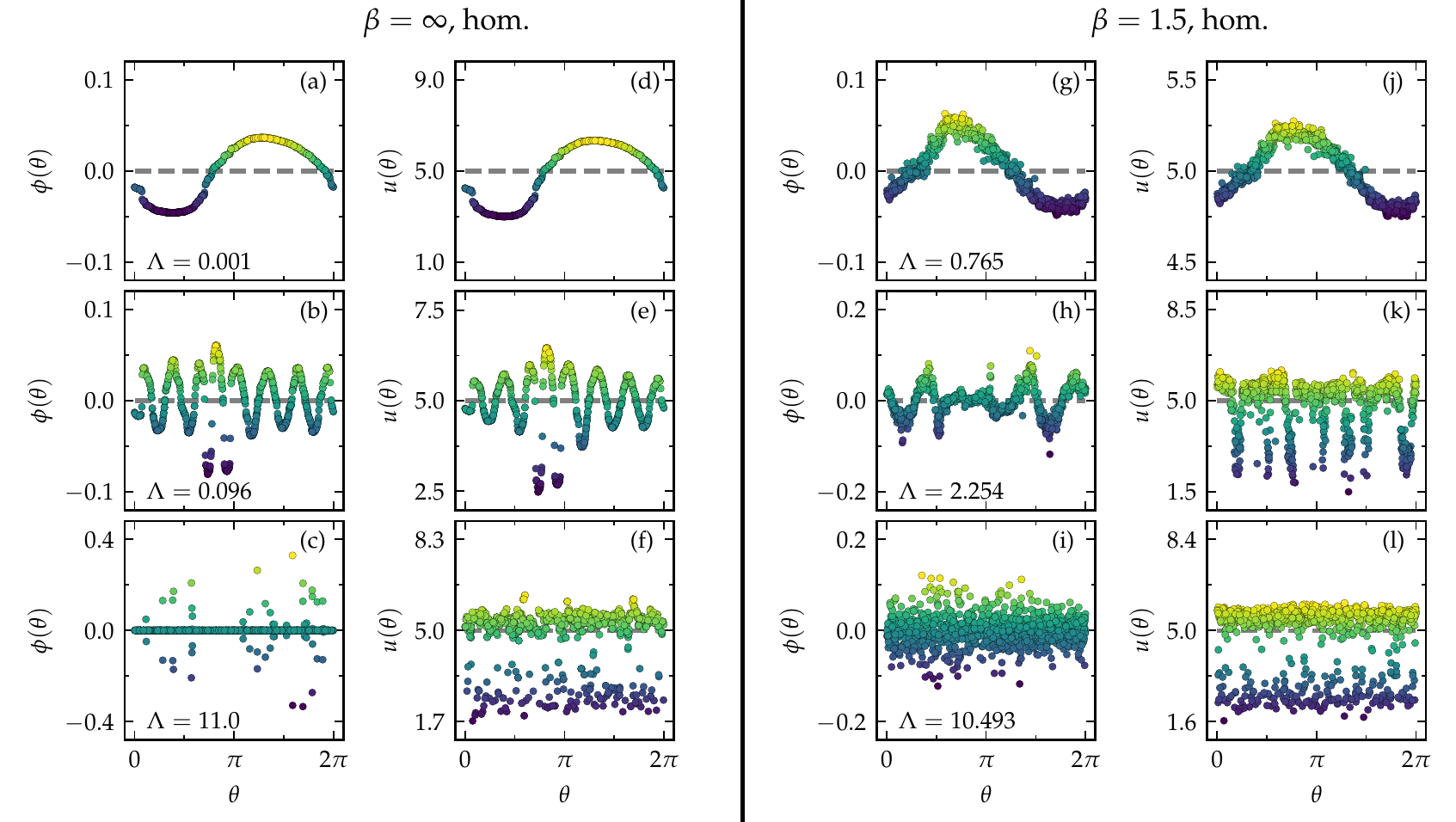}
	\caption{a-c,g-i) Examples of three eigenvectors of the quenched Laplacian matrix with low, medium, and high associated frequencies for a single network generated by the homogeneous ensemble with $N=1000$ and $\langle k \rangle=20$ and $\beta=\infty$ (panels (a-c)) and $\beta=1.5$ (panels (g-i)). d-f,j-l) Concentration of species $U$ in the Mimura-Murray dynamics for the eigenvalues in the column to the left of each panel.} 
	\label{SampledEigen}
\end{figure*}

The analytic solution of Eq.~\eqref{eigen3} can be found in particular cases. One such case is a homogeneous ensemble where all nodes have the same hidden degree, that is $\rho(\kappa)=\delta(\kappa-\langle k \rangle)$ (Notice that this is equivalent to taking the limit $\gamma\rightarrow\infty$), leading to a Poisson degree distribution of average $\langle k \rangle$. In this situation, the eigenvector is just a function of the frequency $\omega$ and Eq.~\eqref{eigen3} becomes
\begin{equation}
\left[ \frac{\hat{\Psi}(\mu \langle k \rangle^2 \omega)}{\hat{\Psi}(0)}-\frac{\langle k \rangle-\Lambda}{\langle k \rangle}\right]\hat{\phi}(\omega)=0.
\end{equation}
Assuming that $\hat{\phi}(\omega) \ne 0$, this equation defines a dispersion relation between the eigenvalue $\Lambda$ and its characteristic frequency $\omega_c$ as a solution of the transcendent equation
\begin{equation}
\frac{\Lambda(\omega_c)}{\langle k \rangle}= \left[ 1-\frac{\hat{\Psi}(\mu \langle k \rangle^2 \omega_c)}{\hat{\Psi}(0)}\right].
\label{transcendent}
\end{equation}
The characteristic frequency $\omega_c$ in Eq.~\eqref{transcendent} is constrained by the boundary conditions and the discretization of nodes in the space. Indeed, since nodes are distributed on the line at density 1, frequencies above $\pi$ (and so wavelengths of the order 1) cannot be detected. Besides, if the system is finite of length $L=N$, frequencies below $2\pi/N$ will have associated wavelengths comparable to the system size and, so, will not be detected either. Therefore, we will look for solutions of Eq.~\eqref{transcendent} in the domain $\omega_c \in[2\pi/N,\pi]$.

It is illustrative to analyze in detail the homogeneous case with $\beta=\infty$, which corresponds to a connection probability being a step function. In this case, function $\hat{\Psi}(z)$ takes the form
\begin{equation}
\hat{\Psi}(z)=2\frac{\sin{z}}{z}.
\end{equation}
and $\mu=1/(2\langle k\rangle)$. Combining these results, the dispersion relation defining $\omega_c$ becomes
\begin{equation}
\frac{\Lambda(\omega_c)}{\langle k \rangle}= \left[1-\frac{2\sin{\frac{\langle k \rangle \omega_c}{2}}}{\langle k \rangle \omega_c} \right].
\label{modelbetainfinity}
\end{equation}

In the case of $\beta=\infty$, nodes connect to nearest neighbors in the line that are below a certain critical distance, so that the average number of such neighbors is $\langle k \rangle$. Therefore, apart from the fluctuations in the number of neighbors, the dispersion relation in Eq.~\eqref{modelbetainfinity} should be equivalent to the dispersion relation of a one dimensional lattice with $\langle k \rangle$ symmetric nearest neighbors, which reads
\begin{equation}
\frac{\Lambda}{\langle k \rangle}=1-\frac{2}{\langle k \rangle}\sum_{n=1}^{\langle k\rangle/2} \cos{n\omega_c}.
\label{lattice1d}
\end{equation}
Figure~\ref{frequency1d} shows the comparison between Eqs.~\eqref{modelbetainfinity} and~\eqref{lattice1d} for $\langle k \rangle=4,8,$ and $12$. Both expressions are very similar and become identical as the connectivity increases, in agreement with the fact that the continuum approximation becomes exact at infinite average degree. 

\begin{figure}[t]
\centering
\includegraphics[width=\columnwidth]{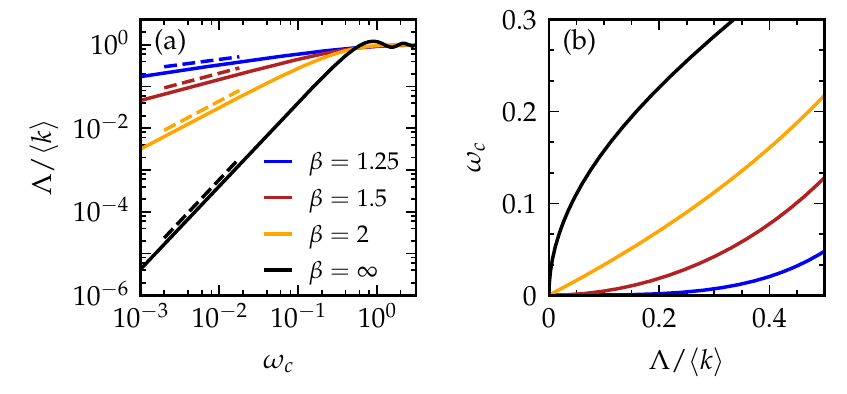}
\caption{\label{frequency1d3}Characteristic frequency for the model for $\langle k \rangle=10$ and $\beta=\infty,2,1.5,$ and $1.25$.}
\end{figure}

Despite the good agreement shown in Fig.~\ref{frequency1d}, the comparison with quenched networks generated by the $\mathbb{S}^1/\mathbb{H}^2$ model has to be made with care. Indeed, for homogeneous networks, the annealed approximation completely neglects fluctuations of nodes' degrees. In a quenched network this is far from true, as the degree distribution follows a Poisson distribution of average $\langle k \rangle$. However, the dispersion relations Eqs.~\eqref{modelbetainfinity} and~\eqref{lattice1d} are strongly dependent on the network connectivity, especially in the high frequency domain. Therefore, due to the randomness of nodes' degrees, it is not possible to characterize each eigenvector of the quenched Laplacian by a unique frequency but by a collection of frequencies around a given average.

%the dispersion relation in a quenched network will be a mixture of dispersion relations with different connectivities. \jk{This final part is not so clear to me. } {\color{blue}not clear to me either :-)}

%\begin{figure}[t]
%	\centering
%	\includegraphics[width=\columnwidth]{EigVec_b15_homogeneous_T.pdf}
%	\caption{a,c,e) Examples of three eigenvectors with low, medium, and high associated frequencies for a single network generated by the homogeneous ensemble with $N=1000$, $\beta=1.5$, and $\langle k \rangle=20$. b,d,f) Concentration of species $U$ in the Mimura-Murray dynamics for the eigenvalues of the left column.} 
%	\label{SampledEigen2}
%\end{figure}

We cannot talk then of a deterministic dispersion relation but a fuzzy one where different frequencies may coexist in the same eigenvector. This will destroy any possible high frequency periodic pattern, leaving only low frequency eigenvectors (with $\Lambda \ll \langle k \rangle$) visible. The reason is that low frequency modes correspond to long wavelengths, so that local fluctuations of degrees become irrelevant. This is clearly visible in Fig.~\ref{SampledEigen}a-c, showing three different eigenvectors of the annealed Laplacian matrix corresponding to low, medium, and high frequencies of a graph generated by the  homogeneous ensemble. The periodic pattern is very clear in the case of low eigenvalue with very low frequency. In the medium frequency case, we can still identify a periodic behavior, although distorted by noise. In the case of high frequency, periodic behavior is totally absent. We also run the Mimura-Murray dynamics on the same network selecting parameters' values such that the same eigenvalues used in Fig.~\ref{SampledEigen}a-c become unstable (see  Appendix~\ref{appendix_dynamics} for technical details of the dynamics). Fig.~\ref{SampledEigen}d-f show the concentration of species $U$ in the three cases. We observe that the patterns in the concentration of species $U$ closely follow the patterns of the corresponding eigenvectors, justifying the relevance of the structure of eigenvectors for dynamical processes on networks.

Based on these observations, we conclude that the annealed approximation provides good results in quenched networks for small eigenvalues such that $\Lambda \ll \langle k \rangle$, which corresponds to low frequencies, that is, $\mu \langle k \rangle^2 \omega_c \ll 1$. For arbitrary values of $\beta$, we can then take the low frequency limit in Eq.~\eqref{transcendent} to derive the relation between $\omega_c$ and $\Lambda$. Using the definition of the Fourier transform, it is easy to see that for small $z$
\begin{equation}
1-\frac{\hat{\Psi}(z)}{\hat{\Psi}(0)} \sim
\left\{
\begin{array}{lr}
 z^{\beta-1}& \beta \le 3\\
 z^2 & \beta > 3
 \end{array}
 \right..
\end{equation}
Figure~\ref{frequency1d3}a shows the low frequency limit of the dispersion relation computed numerically for different values of $\beta$, which corroborates this scaling behavior. Using this result, the characteristic frequency scales as
\begin{equation}
\omega_c \sim 
\left\{
\begin{array}{lr}
\frac{1}{\langle k \rangle}\left[\frac{\Lambda}{\langle k \rangle}\right]^{\frac{1}{\beta-1}} & \beta \le 3\\[0.5cm]
\frac{1}{\langle k \rangle}\sqrt{\frac{\Lambda}{\langle k \rangle}} & \beta > 3.
\end{array}
\right.
\label{eq:omegacbeta}
\end{equation}
Thus, low values of $\beta$ --increasing the number of long-range links-- result in very few eigenvectors with very low frequency and in the limit of $\beta \rightarrow 1$ the fraction of periodic eigenvectors vanishes, as shown in Fig.~\ref{frequency1d3}b.

\begin{figure}[t]
	\centering
		\includegraphics[width=\columnwidth]{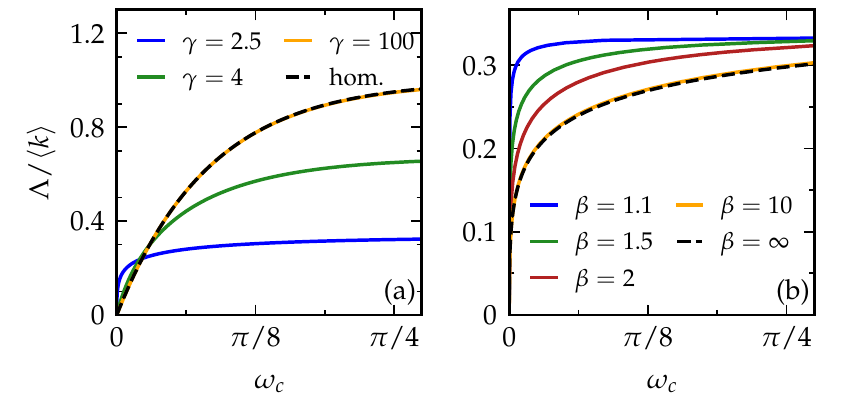}
	\caption{(a) Dispersion relation obtained with Gaussian quadrature for $\beta=2$ for various $\gamma$'s and (b) the dispersion relation for $\gamma=2.5$ and varying $\beta$. The average degree is $\langle k\rangle = 12$ in both panels. }
	\label{FIG:TheoryDispRel}
\end{figure}

Fig.~\ref{SampledEigen}g-l show the same analysis as in Fig.~\ref{SampledEigen}a-f but for the case $\beta=1.5$, so that networks are deep into the small-world regime, with many long-range connections among distant nodes in the metric space. We observe the same qualitative behavior as in the homogeneous case, except that, as expected, results have more noise. Yet, we can clearly identify geometric patterns both in the eigenvectors and in the steady state of the Mimura-Murray dynamics when the dynamic parameters are properly tuned. We notice, however that, in agreement with our theoretical prediction in Eq.~\eqref{eq:omegacbeta}, eigenvalues with similar associated eigenvector frequencies as those in Fig.~\ref{SampledEigen}a,d are significantly larger.

\subsection{Heterogeneous ensemble}

\begin{figure}[t]
	\centering
	\includegraphics[width=\columnwidth]{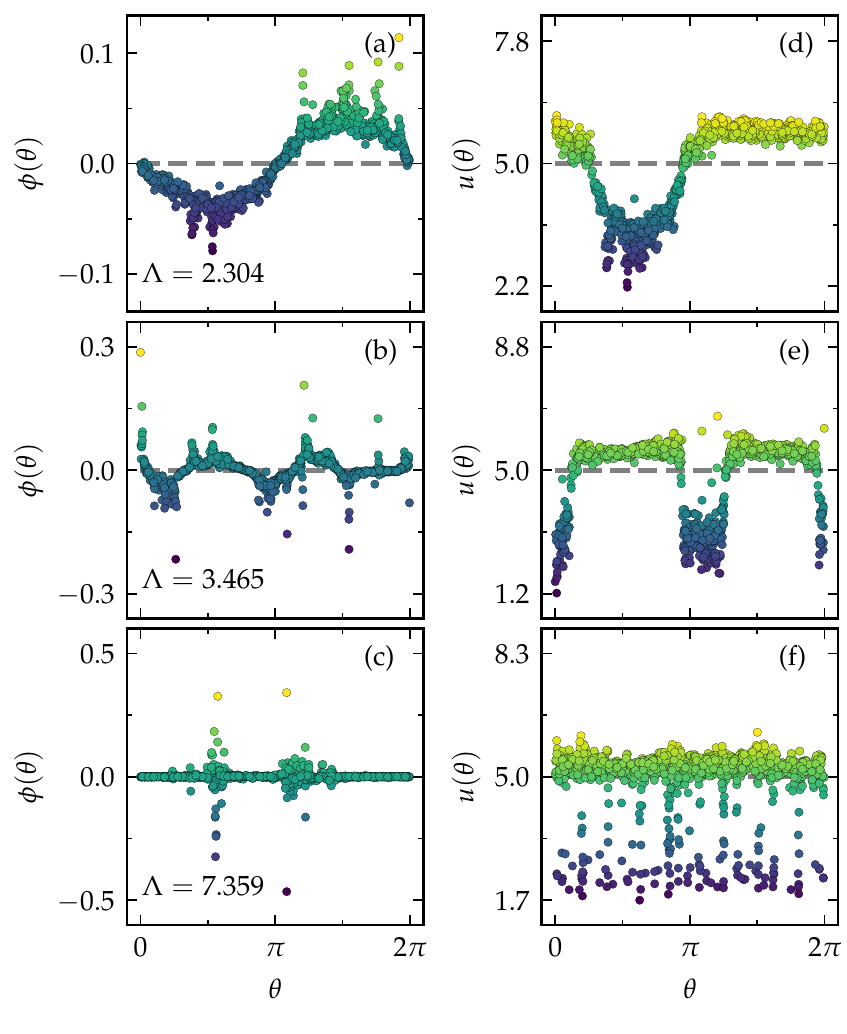}
	\caption{a-c) Examples of three eigenvectors with low, medium, and high associated frequencies for a single network generated by the heterogeneous ensemble with $N=1000$, $\beta=2.5$, $\gamma=2.5,$ and $\langle k \rangle=20$. d-f) Concentration of species $U$ in the Mimura-Murray dynamics for the eigenvalues of the left column.} 
	\label{SampledEigen3}
\end{figure}

In real networks, the degree distribution is typically heterogeneous. Therefore, we have to solve the eigenvalue problem in Eq.~\eqref{eigen3} for a heterogeneous distribution of hidden degrees $\rho(\kappa)$. In particular, we choose a scale-free distribution $\rho(\kappa)\propto\kappa^{-\gamma}$.
For this distribution of hidden degrees, Eq.~\eqref{eigen3} can be rewritten as
\begin{equation}
(\gamma-2)\int_0^1\frac{ z'^{\gamma-3}}{1-\tilde{\Lambda}z'}\hat{\Omega}\left(\frac{\tilde{\omega}}{zz'}\right)\hat{\Theta}(z',\tilde{\omega})\mathrm{d}z'=\hat{\Theta}(z,\tilde{\omega}),
\label{eigen:hetero}
\end{equation}
where we have redefined variables $(\kappa,\Lambda,\omega)$ as 
\begin{equation}
z\equiv \frac{\kappa_0}{\kappa}\; ,\;  \tilde{\Lambda} \equiv \frac{\Lambda}{\kappa_0} \; \; , \; \; \tilde{\omega}\equiv \mu \kappa_0^2 \omega 
\end{equation}
and functions as
\begin{equation}
\hat{\Omega}(x) \equiv \frac{\hat{\Psi}(x)}{\hat{\Psi}(0)} \; \; , \; \; \hat{\Theta}(z,\tilde{\omega})\equiv (1-\tilde{\Lambda}z)\hat{\phi}\left(\frac{\kappa_0}{z},\frac{\tilde{\omega}}{\mu \kappa_0^2}\right).
\end{equation}
When $\tilde{\Lambda}<1$, the singularity of the kernel in the integral of Eq.~\eqref{eigen:hetero} falls outside the domain of integration. In this case, Eq.~\eqref{eigen:hetero} can be solved numerically by Gaussian quadrature as the solution of the system of equations
\begin{equation}
\sum_{j=1}^n  (\gamma-2) \frac{ w_j z_j^{\gamma-3}}{1-\tilde{\Lambda}z_j}\hat{\Omega}\left(\frac{\tilde{\omega}}{z_iz_j}\right)\hat{\Theta}(z_j,\tilde{\omega})=\hat{\Theta}(z_i,\tilde{\omega}),
\label{eigen:quadrature}
\end{equation}
where $z_i$, with $i=1,\cdots,n$, are the zeros of the orthogonal polynomials used in the quadrature, $w_i$ their associated weights, and $n \gg 1$ the number of points within the domain of integration used to evaluate the integral.

Equation~\eqref{eigen:quadrature} defines a homogeneous system of linear equations with kernel matrix $\mathbb{K}(\Lambda,\omega)$ given by
\begin{equation}
K_{ij} \equiv (\gamma-2) \frac{ w_j z_j^{\gamma-3}}{1-\tilde{\Lambda}z_j}\hat{\Omega}\left(\frac{\tilde{\omega}}{z_iz_j}\right)-\delta_{ij}.
\end{equation}
A non-zero solution is found when $\det(\mathbb{K}(\Lambda,\omega))=0$, defining thus the dispersion relation $\Lambda(\omega_c)$. We use this condition to compute numerically the dispersion relation for different values of $\beta$ and $\gamma$. Figure~\ref{FIG:TheoryDispRel}a shows the results for $\beta=2$ and different values of $\gamma$ and Fig.~\ref{FIG:TheoryDispRel}b shows the results for $\gamma=2.5$ and different values of $\beta$. The dispersion relations are qualitatively similar to the homogeneous case with $\beta \le 2$. The main difference appears in the asymptotic behavior of $\Lambda(\omega_c)$, which approaches $\kappa_0$ at high frequencies. Given the relation between the average degree and $\kappa_0$, this result implies that the ratio $\Lambda/\langle k \rangle$ approaches $(\gamma-2)/(\gamma-1)$, a value that is below 1. Therefore, the condition $\langle \Lambda \rangle=\langle k \rangle$ implies that the number of eigenvalues with periodic behavior can only account for a small fraction of all eigenvalues, and this fraction vanishes when $\gamma \approx 2$. We conjecture that the remaining eigenvalues, with $\Lambda>\kappa_0$, are those which cannot be characterized by a single frequency.

Figure~\ref{SampledEigen3} shows the same analysis performed in Figs.~\ref{SampledEigen} but for heterogeneous networks with $\gamma=2.5$ and $\beta=2.5$. The low frequency eigenvector is clearly periodic, although we observe some deviations corresponding to low degree nodes. By increasing the eigenvalue, the periodicity is still preserved although the sinusoidal behavior is strongly modified. Finally, periodicity is destroyed at high eigenvalues. This behavior is, again, translated into the steady state of the Mimura-Murray dynamics so that, even in the presence of strong heterogeneity in the degree distribution, we are able to find geometric Turing patterns in the dynamics.

\section{Numerical results on synthetic $\mathbb{S}^1/\mathbb{H}^2$ networks}
\begin{figure*}[t]
	\centering
	\includegraphics[width=\linewidth]{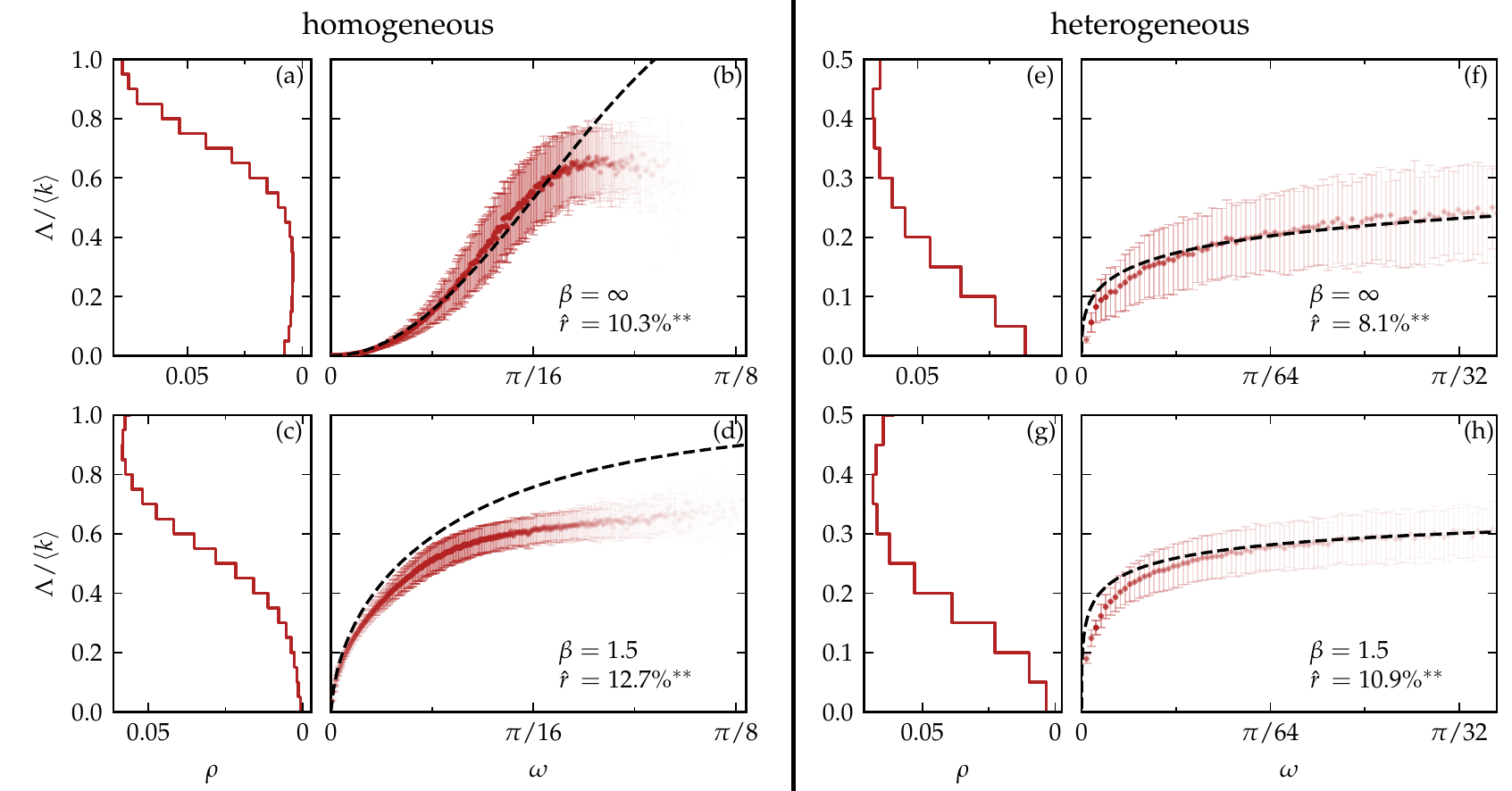}
	\caption{Panels (a,c,e,g): Spectrum of an ensemble of networks generated by the homogeneous (panels (a,c)) and heterogeneous (panels (e,g)) ensembles with $\beta=1.5$ (panels (a,e)) and $\beta=1.5$ (panels (c,g)). For the heterogeneous ensemble $\gamma=2.5$ is used. In all cases the average degree is given by $\langle k \rangle=12$. Panels (b,d,f,h): dispersion relation obtained by applying the Fast Fourier Transform to the eigenvectors and detecting the frequency with highest contribution. The opacity of the points is proportional to total amount of eigenvectors with the frequency corresponding to the point. The more transparent a point is the fewer eigenvectors have that frequency. The black dashed lines are the theoretical predictions given by the annealed approximation. The parameter $\hat{r}$ gives the percentage of eigenvectors classified as periodic by the procedure described in App.~\ref{appendix_fourier}. The notation ${}^{**}$ implies that the results are significant with $p=0.01$.} 
	\label{FIG:Spectrum_3}
\end{figure*}

To confirm numerically our theoretical predictions for the dispersion relations, we need a systematic way of determining when an eigenvector is periodic and, if so, what its dominant frequency is. To do so, we perform a discrete Fourier transform (DFT) on the eigenvectors and apply a statistical $p$-value test with the null-hypothesis being that the Fourier signal is the result of white noise (see Appendix~\ref{appendix_fourier} for further details). We apply this procedure to all eigenvalues and select those with statistically significant periodic signals. For each set of parameters $\beta$ and $\gamma$, we repeat this procedure for a large number of network realizations and average the values of $\Lambda$ obtained corresponding to a given frequency.	

The result of this program is shown in Fig.~\ref{FIG:Spectrum_3}a-d for the homogeneous ensemble with $\beta=\infty$ and $\beta=1.5$, and in Fig.~\ref{FIG:Spectrum_3}e-h for the heterogeneous ensemble with $\gamma=2.5$ and $\beta=\infty$ and $\beta=1.5$. As can be seen, the agreement with the annealed approximation is very good in all cases, with small deviations in the case of $\beta=1.5$. These are understandable because networks generated by the $\mathbb{S}^1/\mathbb{H}^2$ model lose their metric character when $\beta$ approaches one (as clustering goes to zero). Besides, as mentioned in our previous discussion, it is not possible to find periodic behavior at high frequencies as the fluctuations of nodes' degrees induce incoherent signals in the eigenvectors structures. In the examples shown in Fig.~\ref{FIG:Spectrum_3}, the null model combined with the DFT --using a confidence level of two sigmas-- is only able to detect between $8\%$ and $12\%$ of periodic eigenvalues.

\begin{figure*}[t]
	\centering
	\includegraphics[width=\linewidth]{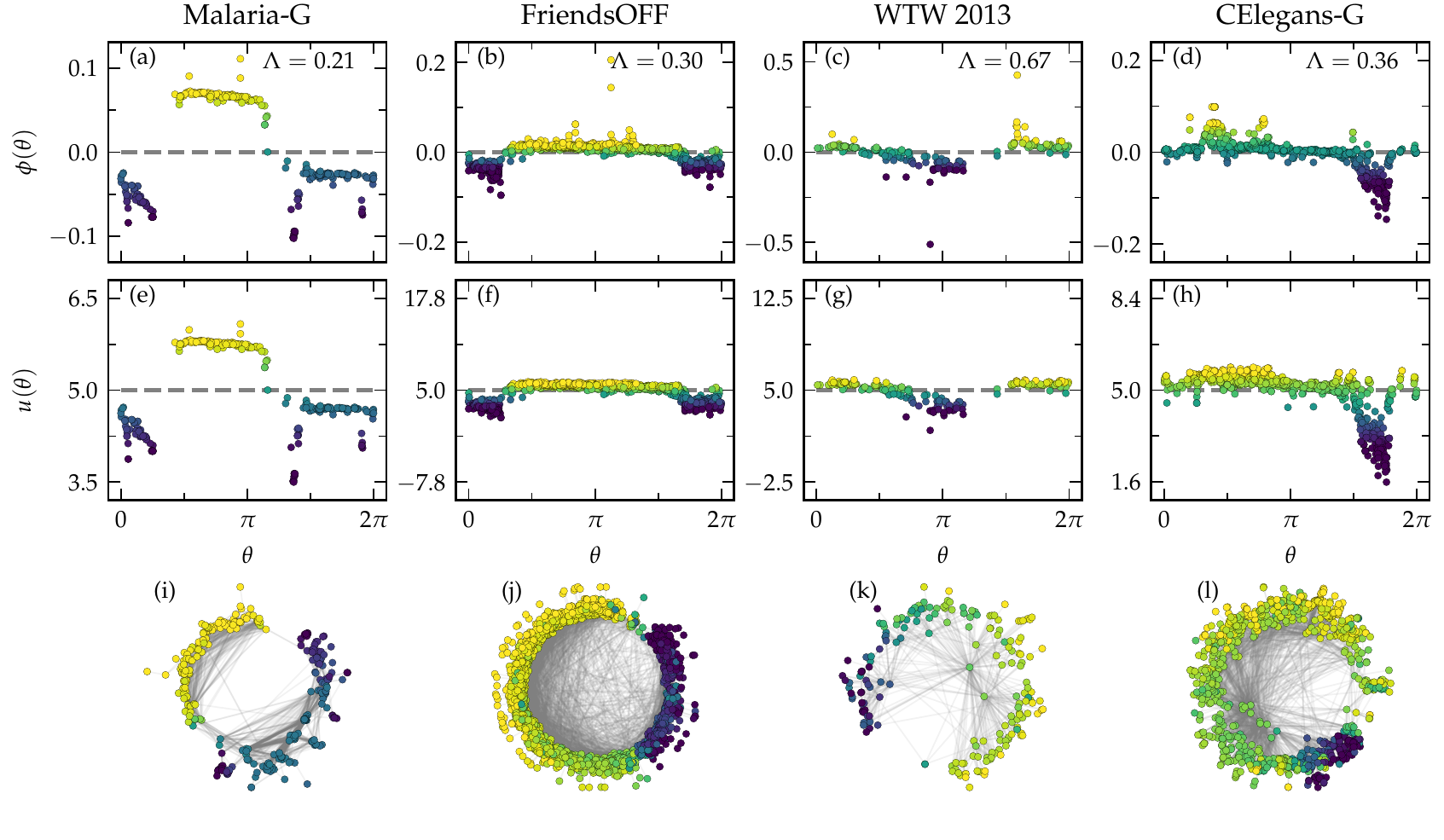}
	\caption{ a-d) Eigenvectors $\phi(\theta)$ corresponding to the eigenvalue $\Lambda$ as denoted in the figure for the real networks analyzed in this study, namely Malaria-G, a network of highly variable genes of the human Malaria parasite, FriendsOFF, an off-line friendship network, WTW 2013, the World-Trade-Web and CElegans-G, a network of genetic interaction of the nematode C. Elegans. e-h) concentration of species $U$ for the Mimura-Murray dynamics setting the parameters to make these eigenvalues unstable. i-l) show the same as e-h) but in the hyperbolic representation of the four networks. } 
	\label{FIG:Real}
\end{figure*}

\section{Real Networks}
\label{real:networks}
To analyze the emergence of Turing patterns in real networks, we first need to find an embedding of the network under study in the hyperbolic plane. This amounts to finding, for each node of the network, its hidden degree $\kappa$ and angular coordinate $\theta$. For this task, we use Mercator~\cite{GarciaPerez2019}, a tool which combines machine learning and maximum likelihood methods to find optimal embeddings congruent with the $\mathbb{S}^1/\mathbb{H}^2$ model. We then run the Mimura-Murray dynamics on the network and look for geometric patterns in the angular similarity space inferred by Mercator. 

Of course, in the case of real networks, we expect less clean results as compared to synthetic graphs generated by the $\mathbb{S}^1/\mathbb{H}^2$ model. Indeed, there are four main factors affecting the emergence of Turing patterns in real networks. First, the degree distribution is generally heterogeneous so that, as in the $\mathbb{S}^1/\mathbb{H}^2$ model, the range of observable frequencies is small. Second, the angular distribution in real networks is not perfectly uniform, with significant fluctuations defining geometric communities which affect the characterization of eigenvectors by a single length-scale. Third, the embedding method is noisy. And fourth, as shown in~\cite{Almagro:2021}, some real networks may be better represented in similarity spaces with dimension higher than $D=1$. Thus, a geometric pattern in a high dimensional space can be distorted in a lower dimensional space, making its detection difficult. Despite these limitations, we expect to find patterns in real networks associated to small non-null eigenvalues.

Figure~\ref{FIG:Real} shows results for four different real networks from different domains. The first is the network of recombinant antigen genes from the human Malaria parasite Plasmodium falciparum, where nodes are var genes encoding for an antigen protein expressed on the surface of the infected red blood cell~\cite{Larremore2013}. These var genes are capable of interchanging bits of genetic information, thus creating a vast amount of slightly different genes that in turn create highly varying antigens that are hard for the human immune system to detect. Two nodes are connected if they share a substring of nucleotides of significant length, indicating that an interchange of genetic material between the two nodes has taken place. The second is an off-line friendship network among students~\cite{Moody2001}, created from a survey where each student was asked to list five male and five female friends. The third network is the World Trade Web, accounting for the import/export interactions among countries in the world~\cite{serrano2003topology,serrano2007patterns,garcia-perez:2016} and the fourth and final network is a genetic network of the nematode C. Elegans, where nodes represent genes and links genetic interactions~\cite{Cho2014}. In Table~\ref{tab:RealNets} we denote the most important network measures for these networks. In the final column we state what percentage of eigenvectors is classified as significantly (with $p=0.01$) periodic by the procedure described in App.~\ref{appendix_fourier}.  Fig.~\ref{FIG:Real}a-d show the structure of the eigenvector corresponding a small eigenvalue (and lower frequency) for the three studied networks. The geometric patterns in the similarity space found by Mercator are very clear, with a wave-length of the order of the system size. The same structure is translated into the dynamics, as shown in Fig.~\ref{FIG:Real}e-h, where the same pattern can be recognized. Fig.~\ref{FIG:Real}i-l show the concentration of species $U$ in the hyperbolic representation of the different networks inferred by Mercator. It is worth stressing how highly non-trivial this result is. Indeed, the Mimura-Murray dynamics generates steady states with correlation lengths of the order of the system size $N$ in small-world networks where any pair of nodes is separated by a very small number of intermediate steps, scaling as $\ln{N}$. Besides, this is achieved despite the fact that neither Mercator nor the Mimura-Murray dynamics use geometric information from the latent space as they only take the bare network topology as input.

\begin{table}[h]
	\caption{Properties of the five real networks studied in Figs.~\ref{fig:results} and \ref{FIG:Real}. The parameter $\hat{r}$ gives the percentage of eigenvectors classified as periodic by the procedure described in App.~\ref{appendix_fourier}. The notation ${}^{**}$ implies that the results are significant with $p=0.01$.}
	\begin{ruledtabular}
		\begin{tabular}{lccccc}
			\hline
			Network         & $N$ & $\langle k\rangle$ & $\beta$ & $\gamma$ & $\hat{r}$\\\hline
			Mouse-C~\cite{Oh2014} & $213$ & $27.9$ & $2.0$ & $3.5$ & $0.9\%^{**}$\\
			Malaria-G~\cite{Larremore2013}  &    $307$ &       $18.3$             &   $2.9$      &   $9$ & $3.9\%^{**}$ \\   
			FriendsOFF~\cite{Moody2001}       &    $2539$ &      $8.2$              &   $1.3$      &    $9$ & $4.5\%^{**}$   \\  
			WTW 2013~\cite{serrano2003topology,serrano2007patterns,garcia-perez:2016}        &    $189$ &     $ 5.8$              &    $1.9$     &  $2.4$  & $5.8\%^{**}$\\   
			CElegans-G~\cite{Cho2014} & $878$ & $7.2$ & $2.6$ & $2.7$ & $2.2\%^{**}$ \\
		\end{tabular}
		\label{tab:RealNets}
	\end{ruledtabular}
\end{table}

\section{Conclusions}

The small-world property, present in the vast majority of complex networks, suggests, {\it a priori}, that networks in the real world are infinite dimensional structures, so that geometry plays a marginal role. Yet, network geometry stands today as the most parsimonious and comprehensive description of this class of systems. However, beyond providing an explanation to the observed topological structure of real networks, it is not clear how network geometry influences the dynamics taking place on them. In this work, we have shown that geometric patterns may emerge in complex networks in a wide class of reaction-diffusion dynamics, as observed also in other types of dynamical processes such as opinion dynamics~\cite{ORTIZ2022112847} and games~\cite{kleineberg2017metric}. Such patterns are already encoded in the structure of the Laplacian matrix as a result of the (hidden) metric structure of real networks. Interestingly, reaction-diffusion processes taking place on such networks are able to sense their geometric nature even when they show extreme disorder induced by heterogeneous degree distributions and a large fraction of long-range links making them small-worlds. 

In this work, we have provided a theoretical framework to understand this phenomenon and to quantify the role that the different topological properties of networks have on the emerging patterns. In addition, significant insight is gained from the fact that Turing patterns can also be detected in real complex networks when the hidden similarity space is properly highlighted. This strongly suggests that there must exists a deep connection between networks' functions and the underlying geometry. Finding such yet unknown connection may shed light on one of the most fundamental questions in network science, namely the connection between the topology and function of complex systems. This is particularly relevant in biological networks, where morphogenesis stands as a fundamental principle behind the organization of living systems. 

\begin{acknowledgments}
We acknowledge support from: Agencia estatal de investigaci\'on project number PID2019-106290GB-C22/AEI/10.13039/501100011033; M. B. acknowledges the ICREA Academia award, funded by the Generalitat de Catalunya; J.~vd~K. acknowledges support from the Secretaria d'Universitats i Recerca de la Generalitat de Catalunya i del Fons Social Europeu.
\end{acknowledgments}

\appendix

\section{Diffusion dynamics on graphs}
\label{appendix:diffusion}

Let us focus on the diffusion of a single species on a simple undirected and globally connected network with adjacency matrix $\mathbb{A} \equiv \{ a_{ij}\} \; \; ; i,j=1,\cdots,N$, and let $n_i(t)$ be the number of particles at node $i$ at time $t$. We assume particles do not interact with each other and that nodes can accumulate an arbitrary number of particles on them. Once at node $i$, a particle jumps away from it at constant rate $\zeta_i$. That is, each particle defines a continuous-time random walk (CTRW)\cite{Montroll:1965hb,Weiss:1994} with dwell time probability density $\psi_i(\tau)=\zeta_i e^{-\zeta_i \tau}$. When a particle jumps from node $i$, it choses one of $i$'s neighbors at random. We can write the following equation for the stochastic evolution of $n_i(t)$ from $t$ to $t+dt$,
\begin{equation}
n_i(t+dt)=n_i(t)+\sum_{j=1}^N \frac{a_{ij}}{k_j} \sum_{l=1}^{n_j(t)} \eta_{j}^l(t,dt)-\sum_{l=1}^{n_i(t)} \eta_{i}^l(t,dt)
\end{equation}
where $\eta_{j}^l(t,dt)$ is a random variable controlling the event that one of the $n_j(t)$ particles sitting at node $j$ jumps to a different node in the interval $(t,t+dt)$. That is
$\eta_{j}^l(t,dt)=1$ with probability $\zeta_j dt$ and $\eta_{j}^l(t,dt)=0$ otherwise. By taking first averages over the random variables $\eta$ and then over $n(t)$, we obtain the equation for the average number of particles at node $i$, $\langle n_i(t) \rangle \equiv u_i(t)$
\begin{equation}
\frac{du_i(t)}{dt}=\sum_{j=1}^N \zeta_j \left( \frac{a_{ij}}{k_j}-\delta_{ij}\right) u_j(t)=-\sum_{j=1}^N \frac{\zeta_j}{k_j} L_{ij}u_j(t),
\label{eq:dif}
\end{equation}
where $L_{ij}$ is the Laplacian matrix~\cite{mieghem_2010}
\begin{equation}
L_{ij}=k_j \delta_{ij}-a_{ij}.
\end{equation}
The steady state solution of Eq.~\eqref{eq:dif} is given by the eigenvector of zero eigenvalue of the Laplacian matrix. This implies that
\begin{equation}
\frac{\zeta_i u_i^{st}}{k_i}=\text{cte}.
\end{equation} 
Therefore, if particles diffuse at the same rate in all nodes --so with $\zeta_i=\zeta \;,\; \forall i$-- then the average concentration at node $i$ is proportional to $k_i$ and only when $\zeta_i \propto k_i$ the average concentration of particles is the same everywhere. We then make the choice $\zeta_i=\epsilon k_i$ and the diffusion equation finally reads
\begin{equation}
\frac{du_i(t)}{dt}=-\epsilon \sum_{j=1}^N L_{ij}u_j(t).
\label{eq:dif2}
\end{equation}
Similarly to the case of diffusion in continuous media, we call $\epsilon$ the diffusion coefficient of the species in the network, even though, unlike diffusion in the continuum, it has dimensions of inverse of time.

\section{Dispersion relation}
\label{appendix_fourier}

To compute the dispersion relation in quenched networks, we generate a large number of them from the $\mathbb{S}^1/\mathbb{H}^2$ model for a given set of parameters $\beta, \gamma$, and $\langle k \rangle$. For each such network, we compute its eigenvectors and eigenvalues sorted in increasing order. We then compute the discrete Fourier transform of the eigenvectors and measure the characteristic frequency as the one corresponding to the highest peak in the Fourier spectrum. To asses whether a peak in the spectrum is statistically significant, we define a null model where the entries of the components of the eigenvector are independently drawn from a normal distribution $\mathcal{N}(\mu,\sigma^2)$ with $\mu=0$ and $\sigma^2=N^{-1}$. The first condition comes from Eq.~\eqref{vanishcomponents} and the second from the fact that eigenvectors are normalized. Denoting $\hat{\phi}_i^{\text{rand}}$ as one of the $N$ entries of the DFT of a random eigenvector from our null model, it can be shown that it satisfies the following distribution function
\begin{equation}
	\text{Prob}\left(|\hat{\phi}^{\text{rand}}_i|^2 < q\right)=1-e^{-q}.
\end{equation}
With this in hand, we can find the value $q$ for which there is a probability $p$ that at least one peak in the white noise Fourier spectrum lies above $q$. We choose a probability $p=10^{-2}$ and take the corresponding value of $q$ to be the minimal value a peak in the Fourier spectrum of one of the eigenvectors of a network generated by the model needs to cross to be considered periodic. Thus,
\begin{equation}
	 p=1- \prod_{i=1}^N \text{Prob}\left(|\hat{\phi}_i|^2<q\right),
\end{equation}
so that
\begin{equation}
q=-\ln\left[1-(1-p)^{\frac{1}{N}}\right].
\end{equation}
For each network, we repeat this procedure for all its eigenvectors, leaving us with a set of pairs $(\Lambda,\omega_c)$ of statistically significant characteristic frequencies, from where we can derive the dispersion relation as an average over many network realizations.

\section{Dynamics on the Network}
\label{appendix_dynamics}

In this paper we use the Mimura-Murray model to perform numerical simulations. In particular, as in Ref.~\citep{Mikhailov:2010}, we choose the parameters as $a = 35$, $b = 16$, $c = 9$ and $d = 2/5$, which gives the fixed point $(\bar{u}, \bar{v}) = (5, 10)$. In this case, the critical value of $\sigma$ above which the Turing instability arises is $\sigma_c=15.5071$. We then set the value to $\sigma=15.6$, slightly above the critical point. Unstable eigenvalues will then lay within the interval $[\Lambda_-,\Lambda_+]$ with
\begin{equation}
	\Lambda_{-}=\frac{1.41026}{\epsilon} \text{  and  } \Lambda_{+}=\frac{1.66667}{\epsilon},
\end{equation}
and the most unstable one given by
\begin{equation}
\Lambda_{\text{max}}=\frac{1.53325}{\epsilon}.
\label{lambdamax}
\end{equation}
To select a given eigenvalue of interest $\Lambda^*$, we set it to $\Lambda^*=\Lambda_{\text{max}}$ and use Eq.~\eqref{lambdamax} to fix the value of $\epsilon$. With all the parameters of the dynamics fixed, we set every node's initial condition to be the stationary one, that is, $(u_i (0), v_i(0)) = (\bar{u}, \bar{v})$ for all $i$ except for a randomly chosen one, $j$, for which $(u_j (0), v_j(0)) = (\bar{u} + \delta, \bar{v} + \delta)$, where $\delta = 10^{-5}$ is a small perturbation. We then leave the system evolve towards its steady state.

\section{The Cheeger constant for the $\mathbb{S}^1/\mathbb{H}^2$ model}
\label{appendix_chegger}

It is worth mentioning an interesting property of the $\mathbb{S}^1/\mathbb{H}^2$ model concerning its spectral gap, defined as the smallest non-null eigenvalue of the Laplacian $\Lambda_2$. Given a graph generated by the $\mathbb{S}^1$ model, $\mathcal{G}_{\mathbb{S}^1}$, Cheeger's inequalities state that 
\begin{equation}
\frac{\left[h(\mathcal{G}_{\mathbb{S}^1})\right]^2}{2k_c}\le \Lambda_2 <2 h(\mathcal{G}_{\mathbb{S}^1}),
\end{equation}
where $h(\mathcal{G}_{\mathbb{S}^1})$ is the isoperimetric (or Cheeger) constant of the graph~\cite{Polya:1951aa,Cheeger:1969aa}. This constant defines the optimal cut of the graph in two disjoint sets of nodes. For any partition $A$ and $B$ in two disjoint sets of sizes $N_A$ and $N_B$, the Chegger constant is defined as the ratio between the number of edges connecting the two sets $E_{AB}$ and the size of the smallest set, minimized over all possible bipartitions of the graph, that is,
\begin{equation}
h(\mathcal{G}_{\mathbb{S}^1})=\min_{A}\left\{ \frac{E_{AB}}{\min(N_{A},N_{B})}\right\}.\label{eq:Cheeger}
\end{equation}

While computing the Cheeger constant is, in general, not possible, in the $\mathbb{S}^1/\mathbb{H}^2$ model we can calculate its scaling behavior with the system size as follows. 
%Let us divide the circle in two equal semicircles and define a binary partition of the graph as the sets of nodes in each semicircle. Then, the average number of edges between a random node in one semicircle to the nodes in the other semicircle scales as $aN^{1-\beta}+bN^{-1}$ and so does the Cheeger constant (see App.~\ref{appendix_chegger} for technical details). Since $\beta>1$, $h(\mathcal{G}_{\mathbb{S}^1})$ approaches zero in the thermodynamic limit and so does the spectral gap. This implies that for very large networks we expect to find eigenvalues arbitrarily close to zero. This is important because, as we see in next sections, small eigenvalues are the ones with more visible and clear patterns.
Assuming that the optimal cut minimizing the Cheeger constant, is made by splitting the unit circle in two continuous regions, in the case of the $\mathbb{S}^1$ model  Eq.~\ref{eq:Cheeger} can be written as follows
\begin{equation}
	h(\mathcal{G}_{\mathbb{S}^1}) = \min_x\bigg(\frac{2\pi}{xN}\int_0^x\int_x^{2\pi}\left(\frac{N}{2\pi}\right)^2\frac{\mathrm{d}\theta_2\mathrm{d}\theta_1}{1+\left(\frac{N\Delta\theta_{12}}{2\pi\hat{\mu}\langle k\rangle^2}\right)^\beta}\bigg),\label{eq:CheegerS1}
\end{equation}
where $\Delta\theta_{12}=\pi-|\pi-|\theta_1-\theta_2||$. Here we defined the two disjoint regions as $A=\{i\,|\,0<\theta_i<x\}$ and $B=\{i\,|\,x<\theta_i<2\pi\}=\mathcal{G}_{\mathbb{S}^1}\backslash A$. As we have chosen $0<x<\pi$, in the model the least amount of nodes reside in $A$ and thus $\min(N_A,N_B)=N_A=xN/(2\pi)$, as seen in Eq.~\ref{eq:CheegerS1}. The second part of this equation represents the amount of links between regions $A$ and $B$, i.e. $E_{AB}$. It can be shown that Eq.~\ref{eq:CheegerS1} can be written as
\begin{equation}
	h(\mathcal{G}_{\mathbb{S}^1}) = \min_x\left(\frac{1}{x}\frac{N}{\pi}\int_0^x\int_{\theta}^{\pi}\frac{\mathrm{d}\theta\mathrm{d}\Delta}{1+\left(\frac{N\Delta}{2\pi\hat{\mu}\langle k\rangle^2}\right)^\beta}\right),
\end{equation}
which in turn can be solved exactly, giving
\begin{equation}
	h(\mathcal{G}_{\mathbb{S}^1}) = 
	\min_x\left(\frac{N}{2\pi}f(N,x)\right),
\end{equation}
where
\vspace{0.3cm}
\begin{widetext}
	\begin{equation}
		f(N,x) = 
		2\pi{}_2F_1\left[
		\begin{array}{c}
			1,\frac{1}{\beta}\\
			1+\frac{1}{\beta}\end{array};-\left(\frac{N}{2\hat{\mu}\langle k\rangle}\right)^\beta\right]+
		x\left(-2{}_2F_1\left[
		\begin{array}{c}
			1,\frac{1}{\beta}\\
			1+\frac{1}{\beta}\end{array};-\left(\frac{Nx}{2\pi\hat{\mu}\langle k\rangle}\right)^\beta\right]+
		{}_2F_1\left[
		\begin{array}{c}
			1,\frac{2}{\beta}\\
			1+\frac{2}{\beta}\end{array};-\left(\frac{Nx}{2\pi\hat{\mu}\langle k\rangle}\right)^\beta\right]\right).
	\end{equation}
\end{widetext}

Here, $_2F_1\left[
\begin{array}{c}
	a,b\\
	c\end{array};z\right]$ is the ordinary hypergeometric function. Note that we use a slightly different form than the standard $_2F_1[a,b;c;z]$ for aesthetic purposes. It can be shown that this scales as  $h(\mathcal{G}_{\mathbb{S}^1})\simeq c_1 N^{1-\beta}+c_2 N^{-1}$, irrespective of the choice of $x$. Since $\beta>1$, this scaling relation implies that $h(\mathcal{G}_{\mathbb{S}^1})$ approaches zero in the thermodynamic limit and so does the spectral gap. This implies that for very large networks we expect to find eigenvalues arbitrarily close to zero. This is important because small eigenvalues are the ones with more visible and clear patterns. This derivation is valid only in the case of a homogeneous hidden degree distribution. However, numerical analysis indicate that the Cheeger constant decays to zero in the case of a heterogeneous distribution as well.

\bibliography{main}

% =================================================================================================
% Acknowledgments
% =================================================================================================
%

\end{document}